\documentclass[preprint,12pt]{elsarticle}

 \usepackage{amssymb,amsfonts,amsthm}
\usepackage[fleqn]{amsmath}
\usepackage{geometry}
\usepackage{graphics,subfigure}
\usepackage{hyperref}
\usepackage{pifont}
\usepackage{natbib}
\usepackage{color}
\usepackage{bm}
\usepackage{ulem}
\usepackage{txfonts}
\usepackage{xcolor}
\usepackage{amssymb}
\usepackage{amsthm}
\usepackage{hyperref}
\usepackage{graphicx}
\usepackage{amsmath,amsfonts,amssymb}
\usepackage{ulem}

\usepackage{soul}

\makeatletter
\def\ps@pprintTitle{%
   \let\@oddhead\@empty
   \let\@evenhead\@empty
   \let\@oddfoot\@empty
   \let\@evenfoot\@oddfoot
}
\makeatother

\hypersetup{
 colorlinks,
 citecolor=Black,
 linkcolor=Black,
 urlcolor=Black}

\begin{document}


\begin{frontmatter}

\title{\textbf{Kirigami-inspired inflatables with programmable shapes}}



\author[a,b,1]{Lishuai Jin\fntext[equal1]}
\author[b,c,1]{Antonio Elia Forte\fntext[equal1]}
 \fntext[equal1]{Authors contributed equally to this work}
\author[b]{Bolei Deng}
\author[d]{Ahmad Rafsanjani}
\author[b,e,f]{Katia Bertoldi\corref{cor1}}

\cortext[cor1]{Corresponding author: bertoldi@seas.harvard.edu}

\address[a]{Department of Mechanics, Tianjin University, 135 Yaguan Road, Jinnan District, Tianjin 300350, China}
\address[b]{John A. Paulson School of Engineering and Applied Sciences, Harvard University, 29 Oxford St., Cambridge, MA 02138}
\address[c]{Department of Electronics, Information and Bioengineering, Politecnico di Milano,  Via Ponzio 34/5, Milan, Italy 20133}
\address[d]{Department of Materials, ETH Z\"{u}rich, 8093 Z\"{u}rich, Switzerland}
\address[e]{Wyss Institute for Biologically Inspired Engineering, 29 Oxford St., Cambridge, MA 02138}
\address[f]{Kavli Institute, Harvard University, Cambridge, MA02138}


\begin{abstract}

Kirigami, the Japanese art of paper cutting, has recently enabled the design of stretchable mechanical metamaterials that can be easily realized by embedding arrays of periodic cuts into an elastic sheet. Here, we exploit kirigami principles to design inflatables that can mimic target shapes upon pressurization. Our system comprises a kirigami sheet embedded into an unstructured elastomeric membrane. First, we show that the inflated shape can be controlled by tuning the geometric parameters of the kirigami pattern. 
Then, by applying a simple optimization algorithm, we identify the best parameters that enable the kirigami inflatables to transform into a family of target shapes at a given pressure. 
Furthermore, thanks to the tessellated nature of the kirigami, we show that we can selectively manipulate the parameters of the single units to allow the reproduction of features at different scales and ultimately enable a more accurate mimicking of the target.

\end{abstract}

\begin{keyword}
kirigami, programmable inflatables, mechanical metamaterials, shape shifting, inverse design
\end{keyword}

\end{frontmatter}








Very popular among children in the form of party balloons, inflatables have also been employed in science and engineering to enable the design of a variety of systems, including temporary shelters~\cite{fritts1991inflatable,kendall1997inflatable,mcniff1999inflatable}, airbags~\cite{hetrick1953safety,jagger1987air}, soft robots \cite{shepherd2011multigait,morin2012,kim2019bio,Martinez2012,li2017fluid,guan2020novel} and shape-morphing structures ~\cite{pikul2017,Ahlquist2017,konakovic2018, siefert2019b}.
To design  shape changing inflatable structures, 
two main strategies have been pursued. On the one hand, load-bearing inflatable structures have been realized using inextensible membranes \cite{fritts1991inflatable,kendall1997inflatable,mcniff1999inflatable,siefert2019a}.
On the other hand, complex shape changes have been achieved by exploiting  the flexibility of 
stretchable membranes with either optimized  initial deflated geometry \cite{siefert2019b,skouras2012,skouras2014,ilievski2011soft} or embedded reinforced components  \cite{kim2019bio,Martinez2012,polygerinos2015,connolly2015,connolly2017,kim2019,belding2018}.

\begin{figure*}[htbp]
\centering
\includegraphics [width=\linewidth]{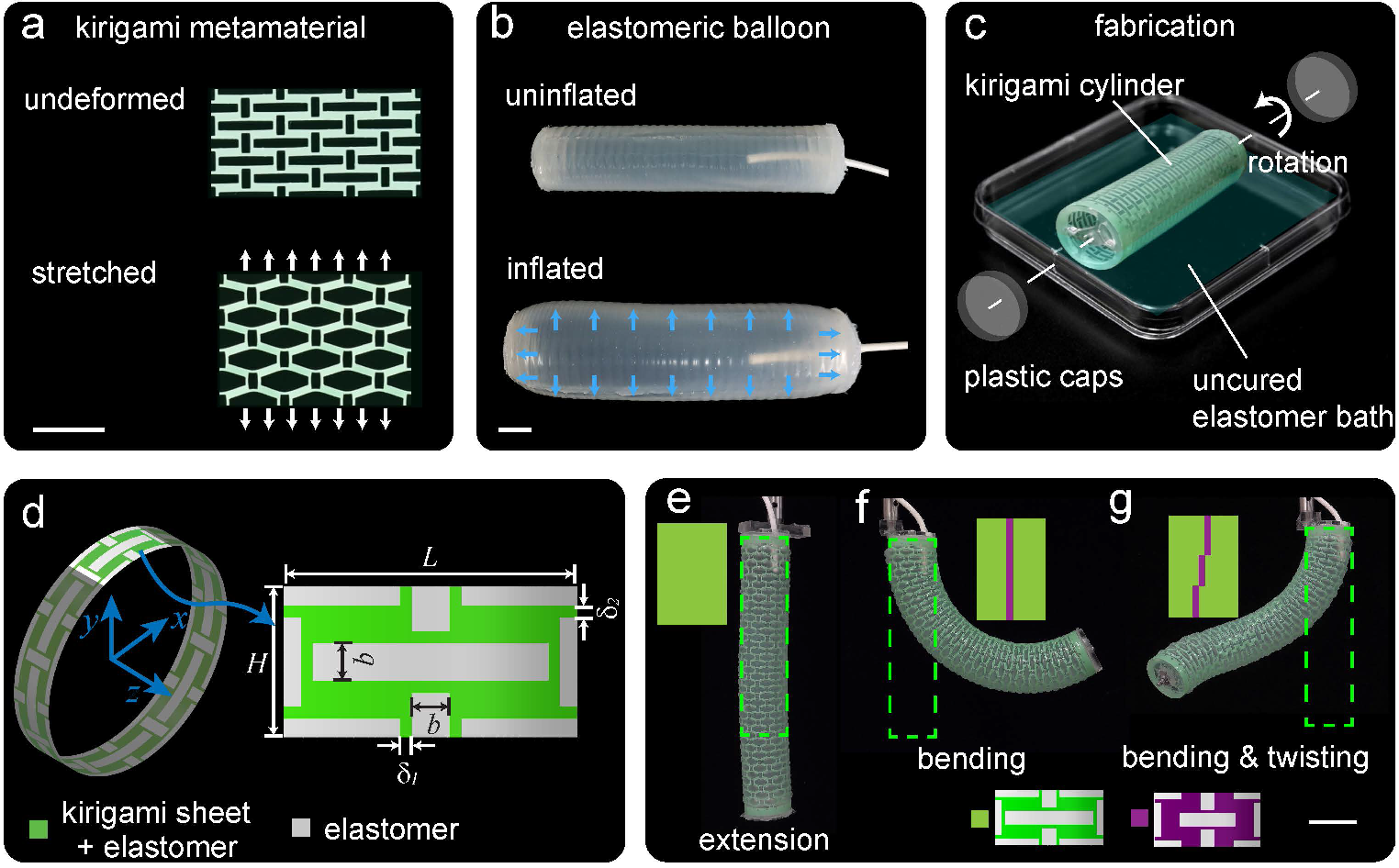}
\caption{Kirigami inflatables. a) A kirigami sheet exhibits large deformation when stretched. b) Deformation of an elastomeric balloon upon inflation. c) Fabrication process. A kirigami shell is rotated in an uncured elastomer bath. Then, the uniformly coated kirigami shell is kept rotating until the elastomer is fully cured. d) Schematic of the kirigami pattern used in this study. e-g) Deformation of  kirigami balloons with $20\times8$ unit cells when subjected to $P$ = 20 kPa. Three design are considered with e) all identical unit cells (with $\delta_1/L$ = 0.03 and $H$ = $L/2$); f)  a single column of unit cells with $\delta_1/L$ = 0.18; g) unit cells with $\delta_1/L$ = 0.18 distributed on different columns.  Scale bars = 30 mm.
}

\label{fig_fabrication}
\end{figure*}

Here, we use kirigami 
as a powerful tool to realize shape-shifting structures that can mimic target shapes upon inflation. Kirigami metamaterials, realized by embedding arrays of cuts in elastic sheets,  have recently shown great promise as design platform for flexible devices~\cite{Tang26407,shyu2015kirigami,lamoureux2015dynamic,dias2017kirigami,rafsanjani2017, rafsanjani2018,rafsanjani2019propagation,rafsanjani2019scirob} and morphing structures \cite{celli2018,choi2019,van2019kirigami,an2020programmable}. Their interesting behaviors have been activated using a variety of strategies, including mechanical forces \cite{dias2017kirigami,tang2015design,tang2017programmable}, magnetic fields~\cite{hwang2018tunable}, light~\cite{cheng2019kirigami}, heat~\cite{cui2018origami,liu2019encoding}, pre-stressed substrates~\cite{zhang2015mechanically,zhao2019buckling} and external pneumatic actuators~\cite{rafsanjani2018,rafsanjani2019propagation}. 
Differently, here we introduce a 
kirigami composite that can be used to create airtight inflatables (i.e. kirigami balloons).  
Our system comprises a kirigami plastic sheet (\textbf{Figure \ref{fig_fabrication}}a) embedded into a thin layer of elastomer (Figure \ref{fig_fabrication}b) and can be actuated pneumatically by injection of compressed air into the composite balloon.
We  show that the deformation of such balloons can be guided towards a target shape upon inflation by optimizing the geometry of the kirigami cuts. Remarkably, since we have control of the geometric features for each unit cell, the deformation of the inflatable can be programmed at the "pixel" level. This enables the realization of inflatables that mimic the target shape at different scales when guided by robust algorithms to optimize their design. While a few strategies have been proposed to control the deformation of kirigami \cite{Tang26407,tang2017programmable}, these are lacking an inverse design approach.
On the other hand, optimization strategies have been successfully developed for the design of shape morphing origami structures \cite{dudte2016programming,CALLENS2018241,felton2014method}. Unfortunately, these cannot be employed when  designing  kirigami  structures as their degrees of freedom are different. Regarding kirigami, although a few approaches have been recently proposed  for their optimization
\cite{choi2019,konakovic2016beyond,Chen4511},  these all focused purely on geometry and did not consider  elasticity in the systems. Differently, in our approach we fully account for the elasticity of the material and demonstrate how this results in an enlarged  design space.

To fabricate our kirigami balloons, we first embed a computationally-designed array of cuts into a polyester plastic sheet (Artus Corporation, NJ, with thickness $t\sim 76.2\mu$m, Young's modulus $E =$ 4.33 GPa and Poisson's ratio $\nu=$ 0.4). Although our approach can be applied to any kirigami geometry, we consider a pattern of mutually orthogonal slits of width $b$   (Figure \ref{fig_fabrication}d) since this particular pattern provides a wide range of tunability for the unit cell's Poisson's ratio (Figure S17, Supplementary Information). The selected unit cell has width $L$ and height $H$ and comprises four rectangular domains  connected by hinges of width $\delta_1$ and $\delta_2$ (in the horizontal and vertical direction, respectively). Throughout the study we consider  $L$=12 mm and $\delta_2/L=0.03$ as fixed parameters and tune the mechanical response of the unit cells by varying  $H/L\in[0.5\sim2.0]$ and  $\delta_1/L\in[0.02\sim0.18]$. To turn the kirigami sheet into an inflatable, we firstly roll it into a cylindrical shell and glue acrylic caps to both ends. Then we slowly rotate the kirigami shell in a bath of uncured silicone rubber (Ecoflex$^{TM}$ 00-50, Smooth-On,  with initial shear modulus $\mu=40.5$ kPa) for 20 minutes. This forms a uniform coating with thickness $t\sim 0.5$ mm that embeds the kirigami sheet completely (Figure \ref{fig_fabrication}c; Section S1 and Movie S1, Supporting Information). Once the elastomer is fully cured, we inflate the system by providing pressurized air and record the deformation with a digital camera (SONY EX100V).

To demonstrate the potentials of kirigami inflatables in Figures~\ref{fig_fabrication}e-g we report experimental snapshots for three kirigami balloons comprising $n_z=20$ and $n_\phi=8$ unit cells in the axial and circumferential direction, respectively.  In the first design, all unit cells are identical and characterized by $\delta_1/L=0.03$ and $H/L=0.5$. As one would expect, upon inflation, this structure deforms homogeneously along its soft axis and mostly elongates (Figure~\ref{fig_fabrication}e).  However, by increasing $\delta_1/L$ to 0.18 for a single column of unit cells, we transform the deformation mode from extension to bending and obtain a curved profile upon inflation (Figure~\ref{fig_fabrication}f). Further, thanks to the tessellated nature of the kirigami, we can choose to distribute the unit cells  with $\delta_1/L$= 0.18  on different columns within the structure and achieve a complex coupled bending-twisting deformation  (Figure~\ref{fig_fabrication}g; Movie S2, Supporting Information).The variation of the structures' deformation over multiple loading cycles has also been tested and found negligible (Figure~S6, Supporting Information). As such, these results  highlight not only the flexibility and potential of our approach, but also the richness of the design space. In the remainder of this paper, we combine Finite Element (FE) analyses and optimization to efficiently explore the myriad of possible designs and identify spatially varying distributions of geometric parameters  resulting in target shape changes upon inflation.

\begin{figure}[htbp]
\centering
\includegraphics [width=\linewidth]{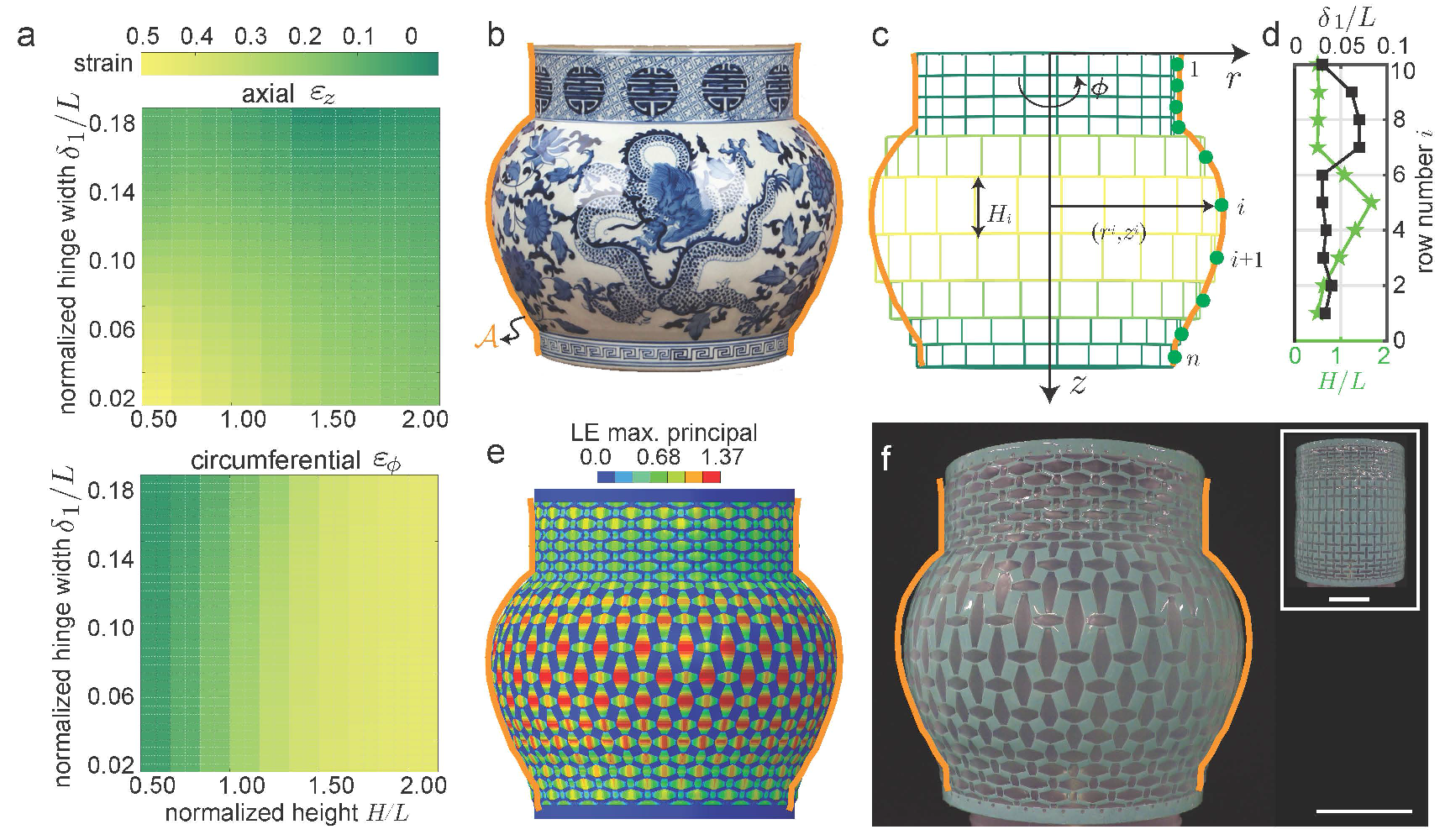}
\caption{Targeting axisymmetric profiles.
a) Evolution of  the axial strain, $\varepsilon_z$, and the circumferential strain, $\varepsilon_\phi$, as a function of  $\delta_1/L$ and  $H/L$ for unit cells with initial curvature $\kappa=2\pi/(n_\phi L)=\pi/(4L)$ subjected to a pressure $P=20$ kPa. 
b) A  jar is selected as target profile. c) Schematic of axisymmetric profile optimization model. d) Optimal design for an inflatable  with $n_z=10$ and $n_\phi=25$ that mimics the jar when subjected to a pressure $P= 6.4$ kPa (Table S1, Supporting Information). e-f) Snapshots of the optimized design after pressurization. The orange line indicates the target profile. Both  e) FE and f) experimental results are shown. Scale bars = 30mm. }
\label{fig_jar}
\end{figure}

We start by focusing on the design of kirigami balloons that mimic target axisymmetric  profiles upon inflation, such as the jar shown in \textbf{Figure \ref{fig_jar}}b. First, we use FE simulations to characterize how local changes in hinge width $\delta_1$ and unit cell height $H$ affect the macroscopic deformation of the system. Since the deformation of our axisymmetric inflatables (for which all unit cells in each row are identical) can be obtained by superimposing  the responses of the individual rows (Figure S11, Supporting Information), we simulate a single unit cell with suitable boundary conditions applied on its edges (Section S3.1, Supporting Information). In Figure~\ref{fig_jar}a we report the numerical evolution of the homogenized axial ($\varepsilon_z$) and circumferential ($\varepsilon_\phi$) strains as a function of $H/L$ and $\delta_1/L$  for unit cells with initial curvature $\kappa=2\pi/(n_\phi L)=\pi/(4L)$  subjected to a pressure $P=20$ kPa. The contour plots indicate  that  $\varepsilon_z$ is inversely proportional to both $\delta_1/L$ and $H/L$, whereas $\varepsilon_\phi$ is mainly affected  by  $H/L$ and increases monotonically  as $H/L$ becomes larger. It is worth noticing that although these results were obtained with a fixed number of unit cells along the circumference, 
we can show that they also describe the deformation of unit cells with arbitrary curvature $\kappa$ subjected to a normalized pressure $\overline{P}=P/\kappa$ which equals to $\overline{P}$=305.6 kPa$\cdot$mm for our selected parameters. In fact, unit cells with same $\delta_1$ and $H$ but different curvature $\kappa$ experience the same state of deformation if subjected to the same normalized pressure $\overline{P}$ (Section S3.3, Supporting Information).

Once we understand how the geometrical features affect the deformation of the unit cells upon inflation, we can search for arrangements that minimize the mismatch between the shape of the
kirigami balloon inflated  at a normalized pressure $\overline{P}=305.6$ kPa$\cdot$mm    and a target surface of revolution defined by a  profile $\mathcal{A}$ (Figure \ref{fig_jar}b). To identify the optimal height of the $i$-th row of unit cells, $H^i$, and the corresponding ligament width, $\delta_1^i$, we minimize
\begin{equation}
\label{radi_opti}
    \mathcal{Z}=\underset{\delta_1^i,\,H^i}{\text{arg\,min}} \, \Big\{|z^{n_z} - H_\mathcal{A}| +  \sum_i^{n_z} d\left[(r^{i},z^{i}),\,\mathcal{A}\right]\Big\},
\end{equation}
where $H_\mathcal{A}$ is the total height of the target profile and $d\left[\mathbf{x},\,\mathcal{A}\right]$ represents the distance between a point with coordinates $\mathbf{x}$ and the closest  point on the target profile \cite{D'Errico2013}. Moreover, $r^i$ and $z^i$ denote the radial and axial coordinates of the center point of the $i$-th row of unit cells in the inflated configuration, which are given by 
\begin{equation}
    r^i = \frac{n_\phi L}{2\pi} (1+\varepsilon_{\phi}^i),
\end{equation}
and
\begin{equation}\label{zi}
    z^i = \frac{H^i(1+\varepsilon_{z}^i)}{2} + \sum_{j=1}^{i-1} H^j(1+\varepsilon_{z}^j).
\end{equation}
Note that  $\varepsilon_{\phi}^i$ and $\varepsilon_{z}^i$ are the homogenized circumferential and axial strain the unit cells undergo in the $i$-th row upon inflation, which,  for each evaluation  of the objective function, are obtained by linearly interpolating the FE results of Figure~\ref{fig_jar}a. Finally, we solve the  optimization problem described by Eqs. (\ref{radi_opti})-(\ref{zi}) using  a Matlab implementation of the Nelder-Mead simplex algorithm  with bounds  applied to all variables (i.e. we impose $H^i/L\in[0.5\sim2.0]$ and  $\delta_1^i/L\in[0.02\sim0.18]$)~\cite{D'Errico2012}.

In Figure \ref{fig_jar}c we show an inflatable kirigami designed with $n_z=10$ and $n_\phi=25$ that mimics the jar of Figure \ref{fig_jar}b when subjected to a pressure $P=6.4$~kPa (resulting in $\overline{P}=305.6$ kPa$\cdot$mm). Note that the parameters $n_z$ and $n_\phi$ define the resolution of the programmed deformed shape. We explore different combinations of $n_z$ and $n_\phi$ (Figure S12, Supporting Information), and choose the one that provide a small mismatch from the target shape without complicating the fabrication process.  As shown in Figures~\ref{fig_jar}d, the  solution identified by the algorithm for $n_z=10$ and $n_\phi=25$ comprises unit cells with large height $H$ between the third and sixth rows (to maximize the radial expansion) and with large $\delta_1$ in the seventh, eighth and ninth row (to minimize both  axial and circumferential strains). We find that, using the optimized set of parameters, both the FE simulations  and the physical samples closely mimic the target shape upon inflation (Figures~\ref{fig_jar}e and f; Movie S3, Supporting Information), confirming the validity of our approach.

\begin{figure*}[htbp]
\centering
\includegraphics [width=\linewidth]{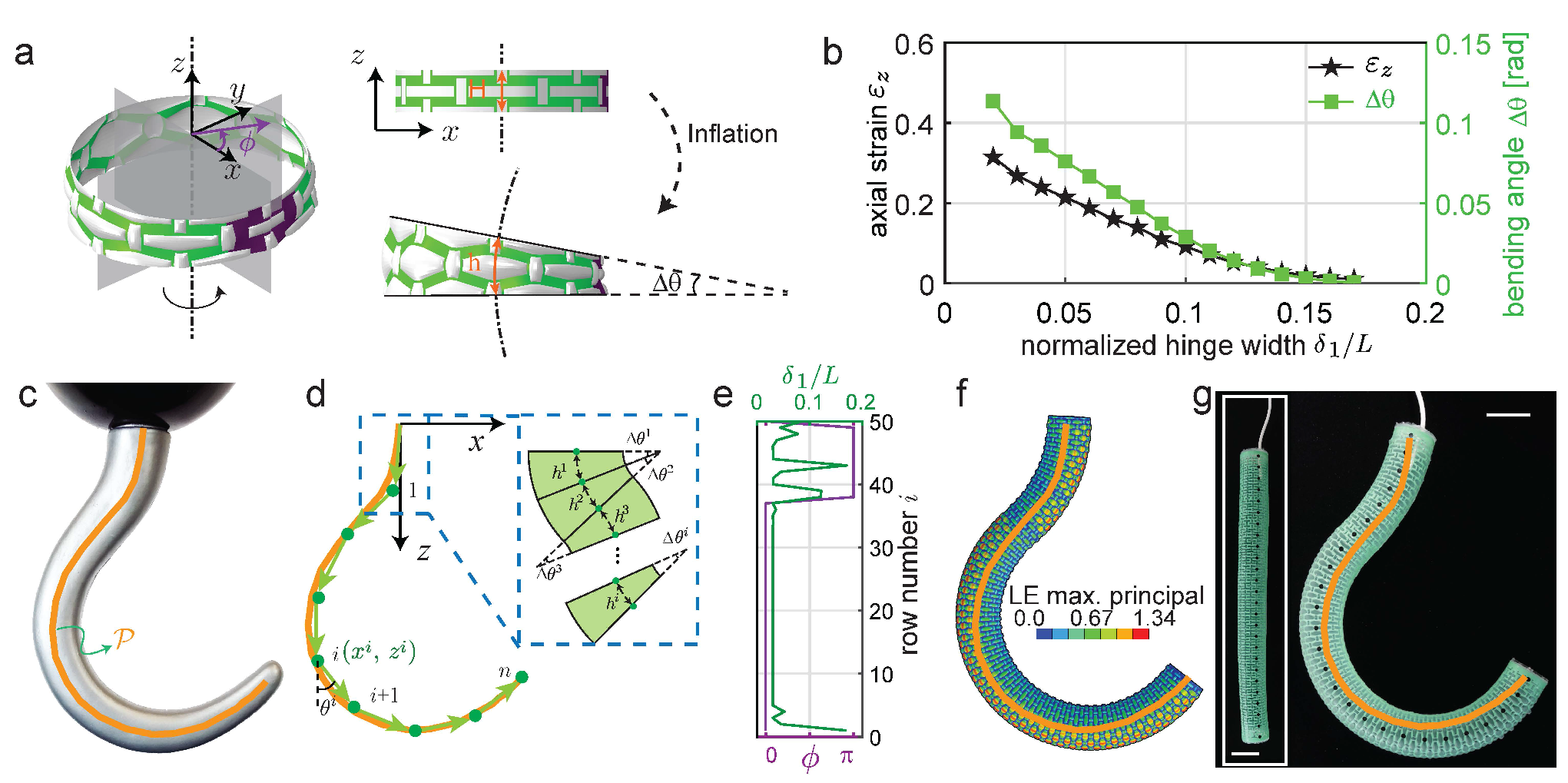}
\caption{ Targeting curvilinear paths.
a) Schematic of a kirigami ring comprising  one unit cell with $H/L$ = 0.5 and $\delta_1/L$ = 0.18 (shown in purple) and $n_\phi$-1 = 7 unit cells with $H/L$ = 0.5 and $\delta_1/L$ $<$ 0.18 (shown in green). The deformation of the ring can be characterized by the axial strain $\varepsilon_z$ and the bending angle $\Delta\theta$.
b) Evolution of axial strain $\varepsilon_z$ and bending angle $\Delta\theta$ as a function of the normalized hinge width $\delta_1/L$ for a ring with initial curvature $\kappa$ = $\pi/(4L)$ subjected to a pressure $P$ = 20 kPa.
c) A hook is chosen as target shape. d) Schematic of curvilinear path optimization model. e) Optimized design for an inflatable with $n_z$ = 50 and $n_\phi$ = 8 that mimics the hook when subjected to a pressure $P$ = 20 kPa (Table S2, Supporting Information). f-g) FE and experimental snapshots of the optimized inflatable kirigami structure when subjected to a pressure $P$ = 20 kPa. Scale bar = 30mm. 
}
\label{fig_hook}
\end{figure*}

Next, we demonstrate how to design kirigami balloons that mimic a planar curvilinear path $\mathcal{P}$ upon inflation. A bending deformation requires unit cells with different geometric features to be arranged in the same row of the kirigami pattern. Therefore, guided by the results of Figures \ref{fig_fabrication}f and \ref{fig_jar}a,  we design the $i$-th row of the kirigami to include one unit cell with $H/L=0.5$ and $\delta_1/L$=0.18 (shown in purple in \textbf{Figure \ref{fig_hook}}a) and $(n_\phi-1)$ unit cells  with the same height (i.e. with $H/L=0.5$) and variable $\delta_1^i/L$ (shown in green in Figure \ref{fig_hook}a). However, since the coexistence of different unit cells on the same row of the kirigami causes non-negligible coupling between these units in the circumferential direction, we cannot directly use the results of Figure \ref{fig_jar}a to predict the effect of $\delta_1^i$ on the bending deformation (Figure S13, Supporting Information).  
Instead, we simulate a full ring with $n_\phi=8$ when subjected to $P$=20 kPa (Section S3.2, Supporting Information) 
and extract the axial strain $\varepsilon_z$ and  the bending angle $\Delta\theta$ (Figure \ref{fig_hook}a). In Figure \ref{fig_hook}b we show the evolution  for both $\varepsilon_z$ and  the  $\Delta\theta$ as a function of $\delta_1/L$. The results  indicate that,  as the hinge width $\delta_1$ increases, both $\varepsilon_z$ and $\Delta\theta$ monotonically decrease (i.e. the bending deformation become smaller).

To identify the design of a kirigami balloon that mimics a prescribed planar curvilinear path $\mathcal{P}$ upon inflation (Figure \ref{fig_jar}c), 
we  assume that the final shape of the inflated kirigami structure can be captured by linearly combining the response of  $n_z$ rings. We then determine both the optimal  $\delta_1^i$ for the $i$-th row and the location of the stiffer unit cell (with $\delta_1/L=0.18$) in the ring by using the Melder-Nelson algorithm with bounds~\cite{D'Errico2012}. Specifically, we minimize 
\begin{equation}
\label{bend_opti}
    \mathcal{Z}=\underset{\delta_1^i}{\text{arg\,min}} \, \Big\{\Big|\sum^{n_z}_{i=1}h^i - L_\mathcal{P}\Big| +  \sum_i^{n_z} \text{d}\left[(x^i,z^i),\,\mathcal{P}\right]\Big\},
\end{equation}
where $h^i = H^i (1+\varepsilon_z^i)$ and $L_\mathcal{P}$ is the total length of the curve $\mathcal{P}$. Further,
$x^i$ and $z^i$ denote the position of the center line at the bottom of the $i$-th ring which can be expressed as 
\begin{equation}
\label{eq_coords}
    x^i = \sum_{j=1}^i h^j\sin\theta^j,\, \text{ and } z^i = \sum_{j=1}^i  h^j \cos\theta^j
\end{equation}
where
\begin{equation}
    \label{eq_Theta}
    \theta^j = \frac{\cos \phi^j\Delta\theta^j}{2}+ \sum_{k=1}^{j-1}\cos \phi^k\Delta\theta^k.
\end{equation}
Note that the  angle $\phi^i$ points at the location of the stiffer cell within the $i$-th ring (Figure \ref{fig_hook}a). It is worth noticing that in the case of 2D curvilinear paths (as those considered here), this angle can only assume two values: $\phi^i=0$ or $\phi^i=\pi$. In fact, our model outputs $\phi^i=0$ if, for the $i$-th ring, the bending angle $\Delta\theta$ defines a positive curvature (e.g. the third segment in Figure \ref{fig_hook}d) and $\phi^i=\pi$ if defines a negative curvature (like the first and second segments in Figure \ref{fig_hook}d).

In Figure \ref{fig_hook}e we consider  an inflatable design with $n_z=50$ and $n_\phi=8$ that mimics the shape of the hook shown in Figure \ref{fig_hook}c when subjected to a pressure $P=20$ kPa (resulting in $\bar{P}=305.6$ kPa$\cdot$mm). 
As shown in Figures \ref{fig_hook}f and g, using the optimized design, both the FE simulation and the experimental model morph from a cylinder to the target hook path upon inflation (Movie S4, Supporting Information).

\begin{figure*}[htbp]
\centering
\includegraphics [width=\linewidth]{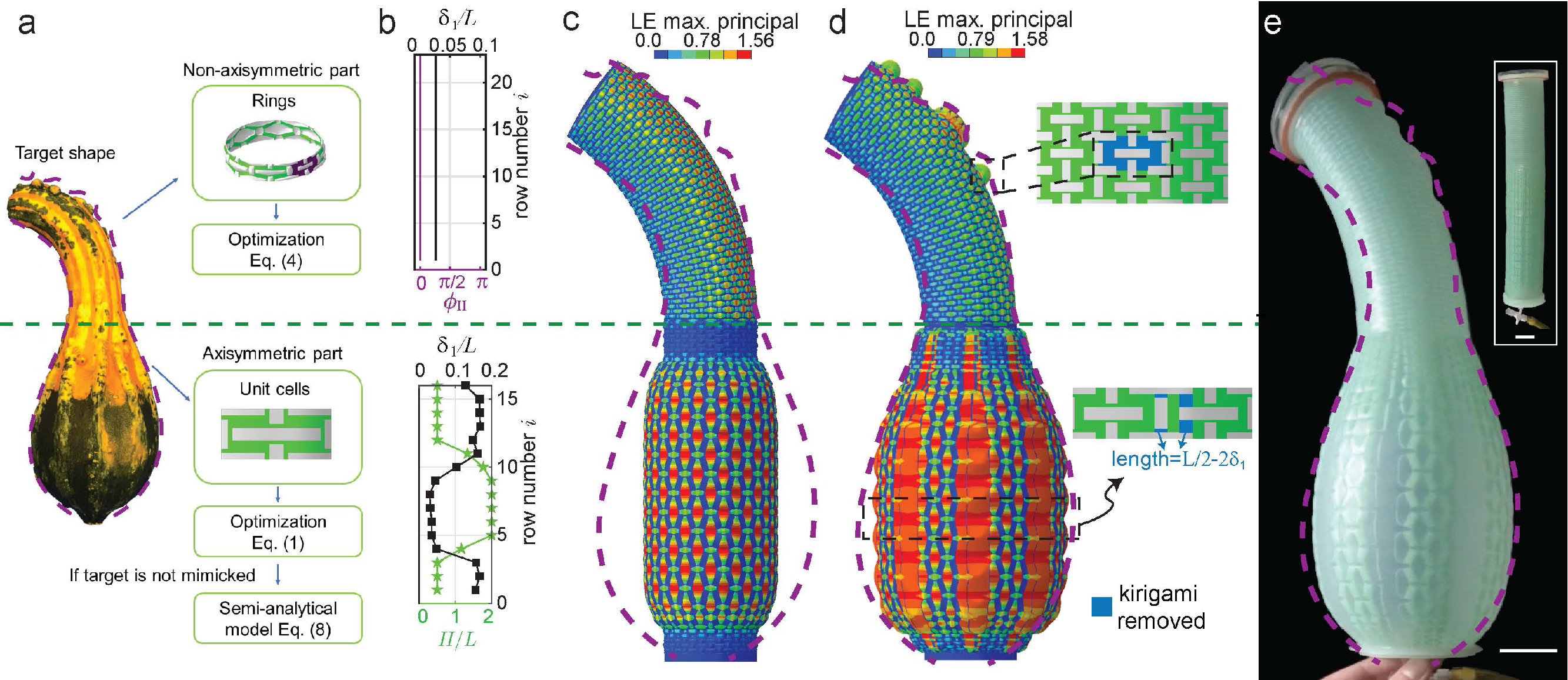}
\caption{Targeting complex shapes. a) A squash is chosen as target shape. The flowchart instructs on the steps to follow in order to optimize for both non-axisymmetric and axisymmetric parts. b) Optimized design for an inflatable with $n_z$=39 and $n_\phi$=16 that mimics the  squash when subjected to a pressure $P$=10 kPa (Table S3 and S4, Supporting Information). The geometric parameters for the top 23 rows are identified using Eq. (\ref{radi_opti}), while those for the bottom 16 rows are obtained using Eq. (\ref{bend_opti}). c) Numerical snapshot of the optimized design after pressurization. The shape of the fruit is not fully captured. d-e) To improve the design we further manipulate the unit cell and remove portions of the kirigami pattern. The bulges can be obtained by removing entire unit cells at the desired location and the circumferential strain in the bottom part can be increased by removing  strips.  Both d) FE  and e) experimental  snapshots of the kirigami inflatable show improved mimicking of the target. Scale bar = 30 mm.}
\label{fig_squash}
\end{figure*}

While in Figures \ref{fig_jar} and \ref{fig_hook} we focused on inflatable that purely expand or bend, the combination of these two classes of deformations enables the mimicking of a multitude of shapes.  As an example, let us consider the squash shown in \textbf{Figure~\ref{fig_squash}}a as target shape. Firstly, we focus on top portion of the fruit, which predominantly bends, and use Eq. (\ref{bend_opti}) to identify the optimal geometric parameters for the corresponding part of the kirigami balloon (Figure \ref{fig_squash}b - top). Secondly, we consider the bottom part of the squash, which follows an axisymmetric profile, and use Eq. (\ref{radi_opti}) to design the corresponding kirigami pattern (Figure \ref{fig_squash}b - bottom). However, the resulting optimized design  does not closely match  the target shape (Figure \ref{fig_squash}c).
Specifically, while the top part of the fruit is successfully reproduced by the optimized inflatable, this fails to mimic the localized bulges near the tip. Moreover, the expansion of the optimized balloon in the lower part is physically limited in the radial direction, resulting in an unsatisfactory transformation.

Nevertheless, we can overcome both limitations by manipulating  the geometrical features of the unit cells even more.  
For example, by removing entire unit cells from the top part of the kirigami pattern (see region highlighted in blue in Figure \ref{fig_squash}d), we are able to obtain localized regions that bulge upon inflation, mimicking the real features of the fruit (Figure \ref{fig_squash}d - top).
Following the same strategy, we can also improve the circumferential stretchability of the bottom part of the structure by selectively removing strips from the kirigami sheet. To determine the width of these sacrificial portions, we first quantify the circumferential strain that a strip of elastomeric material undergoes at a pressure of $P=10$ kPa (resulting in $\bar{P}=305.6$ kPa$\cdot$mm, since in our design $n_\phi=16$). We assume that such strip behaves as an inflated thin elastomeric cylindrical balloon with axial expansion constrained by the kirigami  and   obtain its circumferential strain, $\varepsilon_\phi^e$,  by solving
~\cite{haughton1979bifurcation,fu2010stability} (Section S4, Supporting Information)
\begin{align}
    \label{eq_equilibrium1}
    &P=\frac{t}{r}({\lambda}^e_\phi (1+\varepsilon_z))^{-1}\frac{\partial \hat{W}^e}{\partial {\lambda}^e_\phi},
\end{align}
 where $\lambda^e_\phi=\varepsilon^e_\phi+1$ and $\varepsilon_z$ is the axial strain  of the kirigami (which is provided in Figure \ref{fig_jar}a). Moreover,  $t$ and $r$ denote the thickness and radius of the strip in the undeformed configuration (for our design $r=n_\phi L/(2 \pi)=30.56$ mm and $t=0.5$ mm) and $\hat{W}^e$ is the strain energy function used to captured the response of the rubber (in this study we use a Gent model \cite{gent1996new}). 
Once $\varepsilon_\phi^e$ is obtained, the circumferential strain $\varepsilon_\phi^{tot}$ of a kirigami unit cell with a removed elastomeric strip of width $w_e$  can be estimated as
\begin{equation}
\label{eq_strain}
\varepsilon_\phi^{tot}=\frac{(L-w_e)\varepsilon_\phi+w_e\varepsilon_\phi^e}{L},
\end{equation}
where  $\varepsilon_\phi$ is the circumferential strain of the kirigami unit cell, also provided in Figure \ref{fig_jar}a.

To find the optimum $w^e$ for our balloon, we focus on the kirigami row that is closest to the squash's maximum circumference (row 7th with $H = 24$mm and $\delta_1/L=0.03$). For this specific unit  $\varepsilon_z=0.054$ and $\varepsilon_\phi=0.428$, which results in $\varepsilon_\phi^e=3.59$ according to Eq. (\ref{eq_equilibrium1}).  Further, since we must reach $\varepsilon_\phi^{tot} = 1.094$ in the 7th row upon inflation, we obtain $w^e=2.53$ mm from Eq. (\ref{eq_strain}). Guided by these calculations, we use FE simulations to predict how the response of a kirigami unit cell is affected by the removal of  a kirigami strip of width $w^e=2.53$ mm. We find that the deformation of the inflatable in the axial direction is strictly coupled with the location of the removal within the unit cell (Figure S15a, Supporting Information).
Since our objective is  to achieve the target $\varepsilon_\phi^{tot}$ without compromising  $\varepsilon_z$, we next consider two neighboring unit cells and remove a strip of width $2w^e$ from one, while leaving the other intact.
The results for this case improve considerably (Figure S15b, Supporting Information).  
However, if the width of the elastomeric strip is kept constant in all rows, the inflated balloon fails to match the squash profile as the radial expansion is almost constant along the length  (Figure S16c,  Supporting Information). To further improve the response of our balloon, we choose $w_e$ to vary in each row. Specifically, we assume that $2w_e^i=L/2-2\delta^i_1$ (note that $w^{7}_e=2.64$mm, which is very close to the analytically calculated value), since this enables us to incorporate the information from our optimization algorithm  and fabricate the inflatable by simply removing the hinges highlighted in blue in Figure \ref{fig_squash}d. Results for this final design are shown in Figure \ref{fig_squash}d and e and show that our  design nicely mimic  the target shape  upon inflation -- including the localized bulges on the top part of the bending balloon -- in both the FE model and the physical prototype (Movie S5, Supporting Information). Further, to demonstrate that our approach is general and can be used to mimic a range of shapes, we report an optimized design for a cylindrical structure that morphs into a calabash in the supporting information (Figure S19).

To summarize, in the present work we introduced the concept of kirigami inflatables,  shape morphing systems that combine  a kirigami shell and an elastomeric membrane. We showed that the kirigami shell drives the global deformation of the inflatable and that we can control this deformation by carefully designing its geometric features. We demonstrated this by creating inflatable kirigami balloons that can mimic a variety of axisymmetric shapes and curvilinear trajectories and also capture local features such as bulges.  This multiscale mimicking is enabled by the tessellation nature of the kirigami metamaterial, which allows to easily tweak the local parameters -- or even remove parts of the design -- to boost the deformation locally.
Although our approach enable us to reproduce a variety of targets, there are limitation to the shapes one can mimic. Firstly, the maximum radial expansion of the kirigami balloons upon inflation is limited to 1.43 times the initial radius, using the data set shown in Figure 2a. This limitation can be enlarged by using the removal approach through the semi-analytical model. Additionally, the maximum axial extension upon inflation is 1.46 times of the initial length. It is worth noticing that a bending kirigami balloon present a upper limit on the maximum achievable curvature (e.g. 14.4 $1/m$ for rings with 8 unit cells with L = 12 mm). However, increasing the number of unit cells per unit length of the target provides more feasibility to mimic curvilinear path with larger curvature.
Furthermore, the  kirigami  structure  can  not  mimic  convex  surfaces  in  the  circumferential direction  (e.g.   the  ridges  on  the  squash).   In  principle,  inflatable  structures are not able to form ridges upon inflation without additional constraints (e.g.internal strings or braces).
Lastly, it should be noticed that we only used cylinders as starting deflated shape for our structures.  This limits our approach to the mimicking of shapes within the same ``family", compatibly with the mechanical limitations of the structures.  However, the approach is expandable to other initial shapes,  conditionally to the re-running of the database of solution for the new unit cells and super-cells.
As such, our work provides a new platform for shape morphing devices that could support the design of innovative medical tools, actuators and reconfigurable structures.


\medskip
 \noindent\textbf{Acknowledgements} \par 
 \noindent K.B. acknowledges support from the National Science Foundation under Grants No. DMR-1420570 and DMR-1922321. A.E.F. acknowledges that this project has received funding from the European Union’s Horizon 2020 research and innovation programme under the Marie Skłodowska-Curie grant agreement No 798244.
A.R. acknowledges support from Swiss National Science Foundation through Grant P3P3P2-174326.

\medskip

%



\end{document}